\documentclass[fontsize=20pt,twocolumn,prd,superscriptaddress,longbibliography]{revtex4-1}
\usepackage[utf8]{inputenc}
\usepackage{feynmf}
\usepackage{feyn}
\usepackage{csquotes}
\usepackage{slashed}
 \usepackage{csquotes}

\usepackage{rotating}
\usepackage{adjustbox}
\usepackage{xcolor}
\usepackage{mathtools}

\usepackage{float}
\usepackage{graphicx}

\usepackage{amsmath,tikz}
\usepackage[makeroom]{cancel}
\usepackage{empheq}

\usetikzlibrary{matrix,calc,decorations.markings}
\usepackage{amssymb}
\usepackage{amsfonts}
\usepackage{soul}

\usepackage{indentfirst}
\usepackage{afterpage}
\usepackage[titletoc]{appendix}

\usepackage{dcolumn}
\usepackage[hidelinks]{hyperref}
\hypersetup{colorlinks   = true,linkcolor = teal,urlcolor  = teal,citecolor = blue,}

\usepackage{scalerel}
\usepackage{tikz}
\usetikzlibrary{svg.path}

\definecolor{orcidlogocol}{HTML}{A6CE39}
\tikzset{
  orcidlogo/.pic={
    \fill[orcidlogocol] svg{M256,128c0,70.7-57.3,128-128,128C57.3,256,0,198.7,0,128C0,57.3,57.3,0,128,0C198.7,0,256,57.3,256,128z};
    \fill[white] svg{M86.3,186.2H70.9V79.1h15.4v48.4V186.2z}
                 svg{M108.9,79.1h41.6c39.6,0,57,28.3,57,53.6c0,27.5-21.5,53.6-56.8,53.6h-41.8V79.1z M124.3,172.4h24.5c34.9,0,42.9-26.5,42.9-39.7c0-21.5-13.7-39.7-43.7-39.7h-23.7V172.4z}
                 svg{M88.7,56.8c0,5.5-4.5,10.1-10.1,10.1c-5.6,0-10.1-4.6-10.1-10.1c0-5.6,4.5-10.1,10.1-10.1C84.2,46.7,88.7,51.3,88.7,56.8z};
  }
}

\newcommand\orcid[1]{\href{https://orcid.org/#1}{\mbox{\scalerel*{
\begin{tikzpicture}[yscale=-1,transform shape]
\pic{orcidlogo};
\end{tikzpicture}
}{|}}}}

\newcommand{\GeV}{\; \mathrm{GeV}}
\newcommand{\TeV}{\; \mathrm{TeV}}
\newcommand{\beq}{\begin{equation}}
\newcommand{\eeq}{\end{equation}}
\newcommand{\bea}{\begin{eqnarray}}
\newcommand{\eea}{\end{eqnarray}}

\newcommand{\grav} {\widetilde{G}}

\newcommand\iso[2]{\mbox{${}^{#2}${\rm #1}}}

\def\b1#1{\iso{B}{1#1}}

\makeatletter

\def\slash{\@ifnextchar[{\fmsl@sh}{\fmsl@sh[0mu]}}
\def\fmsl@sh[#1]#2{%
  \mathchoice
    {\@fmsl@sh\displaystyle{#1}{#2}}%
    {\@fmsl@sh\textstyle{#1}{#2}}%
    {\@fmsl@sh\scriptstyle{#1}{#2}}%
    {\@fmsl@sh\scriptscriptstyle{#1}{#2}}}
\def\@fmsl@sh#1#2#3{\m@th\ooalign{$\hfil#1\mkern#2/\hfil$\crcr$#1#3$}}
\makeatother

\def\b{\beta}

\def\d{\delta}

\def\g{\gamma}

\def\m{\mu}
\def\n{\nu}

\def\r{\rho}

\newcommand{\mplanck}{\ensuremath{M_{\text{P}}}}

\def\slash#1{\rlap{\hbox{$\mskip 1 mu /$}}#1}   

\makeatletter 
\def\slash{\@ifnextchar[{\fmsl@sh}{\fmsl@sh[0mu]}} 
\def\fmsl@sh[#1]#2{%
  \mathchoice 
    {\@fmsl@sh\displaystyle{#1}{#2}}%
    {\@fmsl@sh\textstyle{#1}{#2}}%
    {\@fmsl@sh\scriptstyle{#1}{#2}}%
    {\@fmsl@sh\scriptscriptstyle{#1}{#2}}} 
\def\@fmsl@sh#1#2#3{\m@th\ooalign{$\hfil#1\mkern#2/\hfil$\crcr$#1#3$}} 
\makeatother 

\definecolor{darkgreen}{rgb}{0,.5,0}

\newcommand{\sg}{{\tilde g}}
\newcommand{\sq}{{\tilde q}}
\newcommand{\Gr}{  {\grav}}

\newcommand{\nn}{\nonumber \\}

\allowdisplaybreaks
\usepackage{bbold}

\begin{document}
\baselineskip=12.3pt

\title{Gravitino thermal production revisited}      

\author{Helmut~Eberl \orcid{0000-0002-1060-4700}}
\email{helmut.eberl@oeaw.ac.at}
\affiliation{\it Institut f\"ur Hochenergiephysik der \"Osterreichischen Akademie
der Wissenschaften, \\
\it A--1050 Vienna, Austria}
\author{Ioannis~D.~Gialamas \orcid{0000-0002-2957-5276}} 
\email{i.gialamas@phys.uoa.gr}
\affiliation{\it National and Kapodistrian University of Athens, Department of Physics, \\
 Section of Nuclear {\rm \&} Particle Physics,  GR--157 84 Athens, Greece }
 \author{Vassilis~C.~Spanos \orcid{0000-0001-8676-3655}}  
\email{vspanos@phys.uoa.gr}
\affiliation{\it National and Kapodistrian University of Athens, Department of Physics, \\
 Section of Nuclear {\rm \&} Particle Physics,  GR--157 84 Athens, Greece }

\begin{abstract}
We calculate  the gravitino   production  rate,  
computing its one-loop thermal   self-energy.
Gravitino production processes   that do not result through   thermal cuts of its  
  self-energy,   have been  identified and     taken  into account.
Correcting   analytical  errors and   numerical approximations  in the previous calculations,  
we present  our   result.  This deviates  from the latest  estimation   by  almost 10\%. 
More importantly,   we provide  a convenient    formula, for calculating the gravitino production rate and
its thermal   abundance, as a function of the reheating temperature of the Universe.
\end{abstract}

\maketitle

\section{  INTRODUCTION}
Extensions of the Standard Model (SM) in the context of  supergravity
provide us with a dark matter (DM) candidate particle the  gravitino,
the superpartner of graviton. It   interacts  purely 
 gravitationally with other particles and thus  naturally escapes  direct or indirect detection, as the current   experimental and observational data 
 on DM searches  suggest. 
  Therefore, the precise knowledge of its cosmological  abundance is  essential    to apply cosmological constraints on these models.  
Gravitinos may  be produced in various ways: (i) nonthermally,  
 from the  inflaton decays~\cite{Kallosh:1999jj,Giudice:1999am,Nilles:2001ry,Kawasaki:2006gs,Endo:2006qk,Ellis:2015jpg,Dudas:2017rpa,Kaneta:2019zgw}, (ii)  much  later around  the big bang nucleosynthesis  epoch, through  
 the decays of   unstable particles~\cite{Kawasaki:2008qe,Kawasaki:2017bqm,Cyburt:2006uv,Cyburt:2012kp}  and (iii) last but not least,  thermally, using a freeze-in production 
 mechanism, as the Universe cools  down from the  
  reheating temperature ($T_\mathrm{reh}$) until now~\cite{Weinberg:1982zq,Ellis:1984eq,Khlopov:1984pf,Moroi:1993mb,Kawasaki:1994af,Moroi:1995fs,Ellis:1995mr,Bolz:1998ek,Bolz:2000fu,Bolz:2000xi,Steffen:2006hw,Pradler:2006qh,Pradler:2006hh,Rychkov:2007uq,Pradler:2007ne,Ellis:2015jpg}. 
  In particular,  assuming    gauge mediated supersymmetry breaking,
 a different production mechanism (freeze-out)  has to be employed~\cite{Giudice:1998bp,Choi:1999xm,Asaka:2000zh,Jedamzik:2005ir}.

The effort in calculating the  thermal gravitino  abundance, using various techniques, methods and approximations, spans over
almost the last four decades. 
Since the gravitinos are mainly thermally produced at very high temperatures, the effective 
theory  of light gravitinos, 
the so-called nonderivative approach, involving only the spin  1/2 goldstino  components, was initially used.  
In this context,  as some of the production amplitudes exhibit infrared (IR) divergences,  they were regularized 
by introducing either a finite thermal gluon mass or an angular cutoff.
Thus,  in~\cite{Ellis:1984eq} the basic $2 \to 2$ gravitino production  processes
    had been   tabulated for the first time and  calculated, see Table~\ref{table:1}. 
This calculation was further improved in~\cite{Moroi:1993mb, Kawasaki:1994af}.  

As the  Braaten, Pisarski, Yuan (BPY) method~\cite{Braaten:1989mz,Braaten:1991dd}
 succeeded in calculating the axion thermal abundance, in~\cite{Ellis:1995mr} it was further applied 
 to the gravitino,  motivated by the fact that  the gravitinolike axion, interacts extremely weakly with the rest of the spectrum. 
Although in~\cite{Bolz:1998ek} the previous  IR regularization technique  was used, in~\cite{Bolz:2000fu,Pradler:2006qh}  the BPY method was employed, 
taking in addition  into account    the contribution of the spin 3/2 pure gravitino components.  

Eventually,   in~\cite{Rychkov:2007uq} the calculation method improved significantly. There  it was  argued that   the basic requirement to apply
 the BPY prescription, i.e.  $g \ll 1$, where  $g$ is the gauge coupling constant,
 is not satisfied in the whole temperature range of the calculation, 
especially  if  $g$ is the strong coupling constant $g_3$. 
 Therefore, the authors  calculated   the   one-loop  thermal gravitino self-energy    numerically      beyond the hard thermal loop  approximation, 
 with the benefit that   
 this  incorporates also  the  $1 \to 2$ processes    besides  the  $2 \to 2$ ones.
  More importantly, it was noticed that the so-called  subtracted part,  i.e.  pieces  of the   $2 \to 2$ squared amplitudes
   for which the self-energy  may  not  account for,  are IR finite. 
 The main numerical result in~\cite{Rychkov:2007uq} on  the gravitino production rate
   differs significantly, almost by a factor of 2, with respect to the previous works~\cite{Pradler:2006qh,Pradler:2007ne}.
 Unfortunately, in~\cite{Rychkov:2007uq} the main analytical results appear to be inadequate. In particular,   in Sec. IV A
 the equations on the self-energy contribution for  the gravitino production  rate,   seem to be   inconsistent even dimensionally. In  addition, the  
 numerical estimation of this self-energy, was computed only  inside the light cone   due to  the limited computation resources of  that time.  
 Furthermore, two  out of the four nonzero  subtracted parts in the corresponding  Table I in~\cite{Rychkov:2007uq},   turn out to be zero.

Motivated by these, in this paper we recalculate   the thermally corrected gravitino self-energy 
and we compute it without numerical approximations  
at the  one-loop  level.  
Eventually, since our final result  for the gravitino production  rate, like in~\cite{Rychkov:2007uq} is  numerical, following~\cite{Ellis:2015jpg} we present an updated  handy    parametrization of this. 
Our   final  result differs  from that shown in~\cite{Rychkov:2007uq} approximately by  10\%.
Moreover, we calculate the gravitino thermal abundance and discuss possible phenomenological consequences.  
\section{The setup} 
As the gravitino is the superpartner of the graviton, 
its interactions are suppressed by the inverse of the reduced Planck mass  $\mplanck=(8\pi\, G)^{-1/2}$. 
Hence, the dominant contributions to its production, in leading order of the gauge group couplings, are  
 processes of the form $a\,b \rightarrow c\, \widetilde{G}$,  where $\widetilde{G}$ stands for gravitino and  $a$, $b$,  $c$ can be 
three superpartners or one superpartner and two SM particles.
The possible processes and the corresponding  squared amplitudes in
$SU(3)_c$ are given in Table \ref{table:1}, where for their denotation by the letters A-J we follow the ``historical" notation of~\cite{Ellis:1984eq}.
\begin{table}[t]
\caption{Squared matrix elements for  gravitino 
production in $SU(3)_c$ in terms of $ g_3^2 \, Y_3/\mplanck ^2$  assuming massless particles, 
$Y_3 =  1 +m^2_{\sg}/(3 m^2_{\tiny 3/2} )$, $C_3 =   24$ and $C'_3 =   48 $. }
\begin{ruledtabular}
\begin{tabular}{cccc}
$X $ & Process &  $ |{\cal M}_{X,\rm full}|^2  $ &  $ | {\cal M}_{X,\rm sub}|^2 $  \\
\hline \\[-3mm]
 A & $g g \to \sg \Gr$ & $4 C_3 ( s + 2 t + 2 t^2/s) $  & $- 2 s C_3 $\\
 B & $   g \sg \to g \Gr$   & $ - 4 C_3 (t + 2 s + 2 s^2/t)$  & $  2 t C_3$  \\
 C & $ \sq g \to q \Gr  $ &  $ 2 s C'_3 $ &  $0 $\\
 D & $ g q \to \sq \Gr $ & $ - 2 t C'_3 $ & $ 0$ \\
 E &  $   \sq q \to g \Gr $  & $- 2 t C'_3 $ & $ 0 $ \\
 F & $\sg \sg \to \sg \Gr $& $8  C_3 (s^2 + t^2 + u^2)^2/(s t u)  $ & $ 0$ \\
 G & $ q \sg \to q \Gr $& $- 4 C'_3 (s + s^2/t) $ &  $0$ \\
 H  & $  \sq \sg \to \sq \Gr $ & $-2 C'_3 (t + 2 s + 2 s^2/t)  $   &  $0 $\\
 I & $q \sq \to \sg \Gr $ & $- 4 C'_3 (t + t^2/s)  $& $ 0$ \\
 J &  $ \sq \sq \to \sg \Gr $ &  $2 C'_3 ( s + 2 t + 2 t^2/s) $ & $ 0$ \\
\end{tabular}
\label{table:1} 
\end{ruledtabular}
\end{table}
In  $SU(3)_c$, the particles $a$, $b$, and $c$ could be gluons $g$, gluinos $\sg$, quarks $q$, or/and squarks $\sq$. 
Analogous processes happen in $SU(2)_L$ or $U(1)_Y$, 
where the gluino  mass $m_\sg \equiv M_3$ becomes  $M_2$ or $M_1$, 
respectively.
In the factor $Y_N \equiv   1 + { m_{\lambda_N}^2 /  (3 m^2_{\tiny 3/2})}$,  
where $m_{\lambda_N} = \{ M_1, M_2, M_3 \}$ and $ m_{\tiny 3/2}$ is the gravitino mass,
the unity is related to the 3/2 gravitino components and the rest  to the 1/2 goldstino part.
For the calculation of the spin 3/2 part in  the  amplitudes,  following~\cite{Rychkov:2007uq}, we have employed the gravitino 
polarization sum
\beq
\Pi^{\tiny 3/2}_{\m \n}(P) =  \sum_{i = \pm 3/2} \Psi^{(i)}_\m \, \overline{\Psi}^{(i)}_\n = -\frac{1}{2} \g_\m \slash{P} \g_\n -  \slash{P} g_{\m\n}  \, ,
\label{eq:polsum}
\eeq
where $\Psi_\m$ is the gravitino spinor and $P$ its momentum. As in~\cite{Rychkov:2007uq}, for the goldstino spin 1/2 part   the 
nonderivative approach is used~\cite{Moroi:1995fs,Pradler:2007ne}. 
The result for the full squared amplitude  has been proved to be the same, either in the derivative or the nonderivative approach~\cite{Lee:1998aw}.

\begin{figure}[t]
\includegraphics[width=0.23\textwidth]{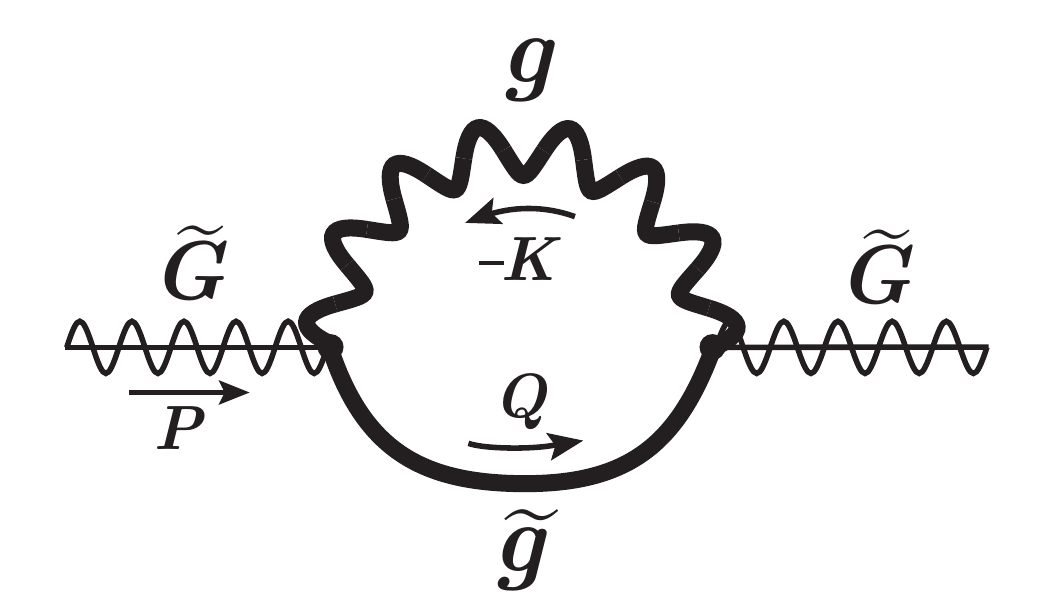}
\caption{The one-loop thermally corrected gravitino self-energy ($D-$graph) for the case of $SU(3)_c$. 
The thick gluon and gluino lines denote resummed thermal propagators.
In our calculation we have taken also into account the equivalent in  $SU(2)_L$ and $U(1)_Y$.}
\label{D_GRAPH}
\end{figure}

The Casimir operators in Table~\ref{table:1} 
are $ C_N =\sum_{a,b,c} |f^{abc}|^2= N(N^2-1)=\lbrace 0,6,24\rbrace$ and
$C_N' =\sum^{\phi}_{ a,i,j} |T^a_{ij}|^2 = \lbrace   11,21,48 \rbrace$,
where $\sum^{\phi}_{ a,i,j}$ denotes the sum over all involved chiral multiplets and  group indices.
$f^{abc}$ and $T^a$ are the group structure constants and generators, respectively.  
Processes A, B and F are not present in  $U(1)_Y$ because $C_1=0$.
 The  masses for the particles  $a$, $b$ and $c$ are assumed to be zero. 
In the third  column of Table \ref{table:1} we present 
for each process the square of the full amplitude, which is the sum of individual amplitudes,
\beq
| {\cal M}_{X,\rm full} |^2 = |{\cal M}_{X,s} + {\cal M}_{X,t} + {\cal M}_{X,u} + {\cal M}_{X,x} |^2 \, ,
\label{Mfull}
\eeq
where the indices $s$, $t$, $u$ indicate the diagrams which are generated by the exchange of a particle in the corresponding channel and 
the index $x$ stands for the diagram  involving a quartic  vertex. The so-called 
$D-$graph, following the 
terminology of~\cite{Rychkov:2007uq},  is  illustrated in Fig.~\ref{D_GRAPH} for the case of the gluino-gluon  loop. 
Its contribution is the sum of the squared amplitudes for the $s$, $t$ and $u$ channel graphs,
\beq
|{\cal M}_{X,D}|^2 = |{\cal M}_{X,s} |^2  +  | {\cal M}_{X,t}|  ^ 2 +  | {\cal M}_{X,u} |^2 \, , 
\label{M2D}
\eeq
plus $1 \to 2$ processes.
This can be understood by  applying  the optical theorem. Hence,   from the imaginary part of the  loop graphs one  computes the sum of the decays ($1 \to 2$) and the scattering amplitudes ($2 \to 2$). In our case, we use resummed  thermal propagators for the gauge boson and gaugino and by applying cutting rules one sees that $D-$graph describes both the scattering amplitudes appearing in~\eqref{M2D} and decay amplitudes.

The subtracted part of the squared amplitudes is the difference between the full amplitudes~\eqref{Mfull} and the amplitudes already included in the $D-$graph~\eqref{M2D}, i.e.  
\beq
|{\cal M}_{X,\rm sub}|^2 =  | {\cal M}_{X,\rm full} |^2 - | { \cal M}_{X,D} |^2 \, .
\label{M2sub}
\eeq
For the processes B,  F,  G, and H, the corresponding amplitudes  are  IR divergent. 
For this reason  we follow  the  more elegant method comprising the separation of 
the total scattering rate into two parts, the subtracted  and the $D-$graph part. It is worth to mention that
for the processes with incoming or/and outgoing gauge bosons, we have checked  explicitly the gauge invariance  for  $|{\cal M}_{X,\rm full}|^2 $.
On the other hand,  we  note that  $|{\cal M}_{X,\rm sub}|^2$ is gauge dependent~\cite{foot3}.  

To sum up, the gravitino production rate $\gamma_{\tiny 3/2}$ consists of three  parts: (i) the subtracted rate $\gamma_{\rm sub}$ (ii) the  $D-$graph contribution  $\gamma_{\rm  D}$ 
and 
(iii) the top Yukawa rate $\gamma_{\rm top}$, 
\beq
\gamma_{\tiny 3/2} = \gamma_{\rm sub} + \gamma_{ \rm D} +\gamma_{\rm top}\,.
\label{gamma}
\eeq
Below,    these three contributions are  discussed  in detail.
\subsection{The subtracted rate}
In the fourth  column of Table \ref{table:1} we present the so-called subtracted part~(\ref{M2sub}), 
which is the sum of the interference terms among the four types of diagrams ($s$, $t$, $u$, $x$),   plus the $x$-diagram squared, for each process.
The subtracted part is nonzero only for the processes A and B.
Note that in~\cite{Rychkov:2007uq} the subtracted part for the processes H and J is also nonzero;  we assume that 
the authors had used  the squark-squark-gluino-goldstino   Feynman rule  as given  in~\cite{Bolz:2000xi}, where a factor $\gamma _5$
  is indeed  missing. In contrast, we are using the correct Feynman rule as given in~\cite{Pradler:2007ne}. 

To calculate the subtracted rate for the processes  
 $a\,b \rightarrow c\, \widetilde{G}$, we use the general form 
 \beq
\begin{aligned}
 \g = \frac{1}{(2\pi)^8} \int &  \frac{{\rm d^3} \mathbf{p}_a}{2E_a}  \, \frac{{\rm d^3} \mathbf{p}_b}{2E_b} \, \frac{{\rm d^3} \mathbf{p}_c}{2E_c} \, 
 \frac{{\rm d^3} \mathbf{p}_{\widetilde{G}}}{2E_{\widetilde{G}}} \, \,  |{\cal M}|^2\,  f_a \, f_b \, (1 \pm f_c)
\\ & \times   \d^4(P_a + P_b - P_c  - P_{\widetilde{G}} )   \, ,
\label{collision_term1}
\end{aligned}
\eeq
where $f_i$ stands for the usual Bose and Fermi statistical densities 
\beq
f_{B|F} = {1 \over e^{E \over T} \mp 1}\, .
\eeq
In the temperature range of interest, all particles but the gravitino  are in thermal equilibrium. For the  gravitino the
statistical factor  $f_{\tiny  {\widetilde{G}}}$ is negligible. 
Thus, $1 -  f_{\tiny {\widetilde {G}}}  \simeq 1$,  as it  is already used in~\eqref{collision_term1}. 
 Furthermore,   backward reactions are neglected. 
In addition,    the simplification $1 \pm f_c  \simeq 1$  is usually applied,
making the analytic calculation of~\eqref{collision_term1}   possible.
In our case there is no such reason. We 
keep the factor  $1 \pm f_c$  and   consequently  
we proceed   calculating   the subtracted rate numerically~\cite{foot1}.

The  contribution of the  processes A and B, for each gauge group,   can be read  
from  Table~\ref{table:1} as
\beq
|{\cal M}_{A , \rm sub}|^2  + |{\cal M}_{B, \rm sub}|^2    =  \frac{g_N^2}{\mplanck ^2}\left(1+ \frac{m^2_{\lambda_N}}{3 m^2_{\tiny 3/2}} \right) C_N (-s + 2 t) \, .
\label{sub_A} 
\eeq
In~\eqref{sub_A}, a factor $1/2$ is already included for the process A due to the two identical incoming 
particles. Substituting ~(\ref{sub_A})   in ~(\ref{collision_term1}), the subtracted rate is obtained as  
\beq
\gamma _{\mathrm{sub}}=\frac{T^6}{\mplanck ^2} \sum_{N = 1}^3 g_N^2 \left(1+ \frac{m^2_{\lambda_N}}{3 m^2_{\tiny 3/2}} \right) C_N \left( -{\cal C}_{\scriptscriptstyle \rm BBF}^s +2\, {\cal C}_{\scriptscriptstyle \rm BFB}^t \right)\,.\label{sub:part}
\eeq
The numerical factors, calculated by using the Cuba library~\cite{Hahn:2004fe}, are $ {\cal C}_{\scriptscriptstyle \rm BBF}^s = 0.25957 \times 10^{-3} $ and 
${\cal C}_{\scriptscriptstyle \rm BFB}^t = -0.13286 \times 10^{-3}\,. $ The subscripts B and F specify if the particles are bosons or fermions, 
espectively, and the superscripts determine if the squared amplitude is proportional to $s$ or $t$. It is easy to see that our result 
for the subtracted part unlike in~\cite{Rychkov:2007uq} is negative. This   is not unphysical,  since the total rate and not the subtracted one is bound to be positive.
\subsection{The $D-$graph contribution}
 As it has been discussed above,   Eq.~\eqref{M2D} describes  
 the relation  between  the $D-$graph  and the sum of the squared amplitudes for  the $s$, $t$, and $u$ channels. 
In the $D-$graph contribution,  we will implement the resummed thermal corrections to the gauge boson  and gaugino  propagators~\cite{foot2}.
Although in Fig.~\ref{D_GRAPH}  the gluino-gluon thermal  loop is displayed, 
  the contributions of all the  gauge groups have been included in our analysis.
 The momentum flow used  to calculate the $D-$graph can be depicted in   Fig.~\ref{D_GRAPH}. That is
    $\Gr(P) \to g(K) + \sg(Q)$,   with 
$ P = (p, p, 0, 0)\,, \,  K = (k_0, k \cos\theta_k, k \sin\theta_k, 0)\,\, \text{and} \,\, Q = (q_0, q \cos\theta_q, q \sin\theta_q, 0)$,
where $\theta_{k,q}$ are the polar angles of the corresponding 3-momenta $\mathbf{k},\mathbf{q} $ in spherical coordinates.

The  non-time-ordered  gravitino self-energy $\Pi^{<}(P)$ can be expressed in terms of the thermally resummed gaugino ${}^*S^<(Q)$ 
and gauge boson ${}^*D^<_{\m\n}(K)$ propagators as~\cite{Bolz:2000fu,Rychkov:2007uq}
\begin{widetext}
\begin{eqnarray}
\Pi^<(P)&=& {1 \over 16 M^2_P} \sum_{N = 1}^3 n_N \left(1 + {m^2_{\lambda_N} \over 3 m^2_{\tiny 3/2}}\right) \int {{\rm d}^4 K \over (2 \pi)^4} \,
{\rm Tr}\left[ \slash{P} [\slash{K}, \g^\m] \, {}^*S^<(Q)  [\slash{K}, \g^\n] \, {}^*D^<_{\m\n}(K) \right]\, ,
\end{eqnarray}
where
\begin{eqnarray}
{}^*S^<(Q) &=&{ f_F(q_0) \over 2} \left[ (\g_0 - {\boldsymbol{\g}} \cdot \mathbf{q}/q) \, \r_+(Q) + (\g_0 + {\boldsymbol{\g}} \cdot \mathbf{q}/q) \, \r_-(Q)\right]\,, \nn
{}^*D^<_{\m\n}(K) & = &  f_B(k_0) \left[ \Pi_{\m\n}^T \,  \r_T(K) + \Pi_{\m\n}^L  {k^2 \over K^2}\, \r_L(K) +\xi {K_\m K_\n \over K^4}\right]\, ,
\label{propagators}
\end{eqnarray}
with  $\xi$ being the gauge parameter, taken $\xi=1$~\cite{foot3}
 in our
calculation and $n_N= \lbrace 1,3,8 \rbrace$. $\Pi_{\m\n}^L$,  $\Pi_{\m\n}^T$, $\r_{L,T}$, and $\r_{\pm}$ are the  
longitudinal, the transverse projectors, and
 the spectral densities 
for the bosons and fermions, respectively. 
 To compute the production rate related to the 
 $D-$graph  $ \gamma_D $,  we will use its definition~\cite{Bellac:2011kqa} 
\begin{equation}
\gamma_D = \int{{\rm d^3 \mathbf{p}} \over 2 p_0 (2 \pi)^3} \,\,  \Pi^<(p)\, 
\label{def_gamma_D_dp}
\end{equation}
and after appropriate  manipulations~\cite{foot4}, we obtain

\bea
\gamma_{\rm D} &=& {1 \over 4(2\pi)^5  \mplanck ^2} \sum_{N = 1}^3 n_N \left(1 + {m^2_{\lambda_N} \over 3 m^2_{\tiny 3/2}}\right)
 \int_0^\infty {\rm d} p \int_{-\infty}^\infty {\rm d } k_0  \int_0^\infty {\rm d} k \int_{|k-p|}^{k+p} {\rm d} q \,\, k \, f_B(k_0)  \, f_F(q_0) \nn 
 &&  \times  \Big[ \r_L  (K) \, \r_- (Q)  \, (p - q)^2  \big( (p + q)^2 - k^2 \big) + \r_L (K) \, \r_+ (Q) \,  (p + q)^2  \big( k^2 - (p - q)^2  \big) \nn  
           &&  + \, \,  \r_T  (K) \, \r_- (Q) \,   \big( k^2 - (p - q)^2  \big)   \Big( (1 + k_0^2 / k^2  \big)   \big( k^2 + (p + q)^2  \big) - 4 k_0 (p + q)  \Big)  \nn 
           && +  \, \, \r_T  (K) \, \r_+ (Q)  \, \big( (p + q)^2 - k^2  \big)  \Big( (1 +  k_0^2 / k^2 )  \big( k^2 + (p - q)^2  \big) - 4 k_0 (p - q)  \Big) \Big] \, ,
\label{gammadgraph}
\eea
where $ q_0=p-k_0\,. $ 
\end{widetext}
The spectral functions $\rho_{L,T}$ and $\rho_{\pm}$ can be found in Eq.~(3.7)  in~\cite{Rychkov:2007uq}. The thermally corrected one-loop  self-energy for
gauge bosons, scalars and fermions that we have used in calculating these spectral functions can be found in~\cite{Weldon:1982aq, Weldon:1982bn,Weldon:1989ys,Weldon:1996kb, Peshier:1998dy,Weldon:1999th}. 
Comparing~\eqref{gammadgraph} with   the corresponding analytical result given in Eqs.  (4.6) and (4.7) in~\cite{Rychkov:2007uq}, 
one can notice that they differ on the overall factor and on the number of independent phase-space integrations.  
Our analytical result has been checked using various  frames for the momenta flow into the loop.
\subsection{The top Yukawa rate}
The production rate resulting from the top-quark Yukawa coupling $\lambda _t$ is given by~\cite{Rychkov:2007uq}
\beq
\gamma_{\mathrm{top}}=\frac{T^6}{\mplanck ^2} \, 72 \,\,  {\cal C}_{\scriptscriptstyle \rm BBF}^s \,  \lambda _t^2  \left(1+ \frac{A_t^2}{3m^2_{\tiny 3/2}} \right)\,,
\label{gamma:top}
\eeq
where $A_t$ is the trilinear stop supersymmetry breaking soft parameter
and ${\cal C}_{\scriptscriptstyle \rm BBF}^s = 0.25957 \times 10^{-3}$.  
 Since this contribution stems from the process  squark-squark $\rightarrow$ Higgsino-gravitino, 
  only the numerical factor ${\cal C}_{\rm BBF}^s$ is involved.
\begin{table}[h]
\caption{\label{table:2} The values of the constants $c_N$ and $k_N$ that parametrize our result~\eqref{gamma:paramtrization} 
for the subtracted and the $D-$graph part. 
Each line  corresponds to the particular gauge group, $U(1)_Y$,  $SU(2)_L$ or $SU(3)_c$.   }
\begin{ruledtabular}
\begin{tabular}{ccc}
Gauge group & $c_N$ & $k_N$\\
\hline 
$U(1)_Y$ & 41.937 & 0.824 \\
$SU(2)_L$ & 68.228 & 1.008 \\
$SU(3)_c$ & 21.067 &  6.878\\
\end{tabular}
\end{ruledtabular}
\end{table}
\section{The parametrization of  the result}
Following~\cite{Ellis:2015jpg} we  parametrize the results~\eqref{sub:part} and~\eqref{gammadgraph} using the gauge couplings $g_1, g_2$ and $g_3\,.$ Thus 
\beq 
\gamma_{\mathrm{sub} }+ \gamma_{\mathrm{D}} =  
 {3  \zeta(3) \over 16  \pi^3 } \, \frac{ T^6}{\mplanck ^2} 
          \sum_{N = 1}^3 c_N \, g_N^2  \left(1 + {m^2_{\lambda_N} \over 3 m^2_{\tiny 3/2}}\right) \ln \left( {k_N \over g_N}\right),
\label{gamma:paramtrization}
\eeq
\begin{figure}[t]
\includegraphics[width=0.483\textwidth]{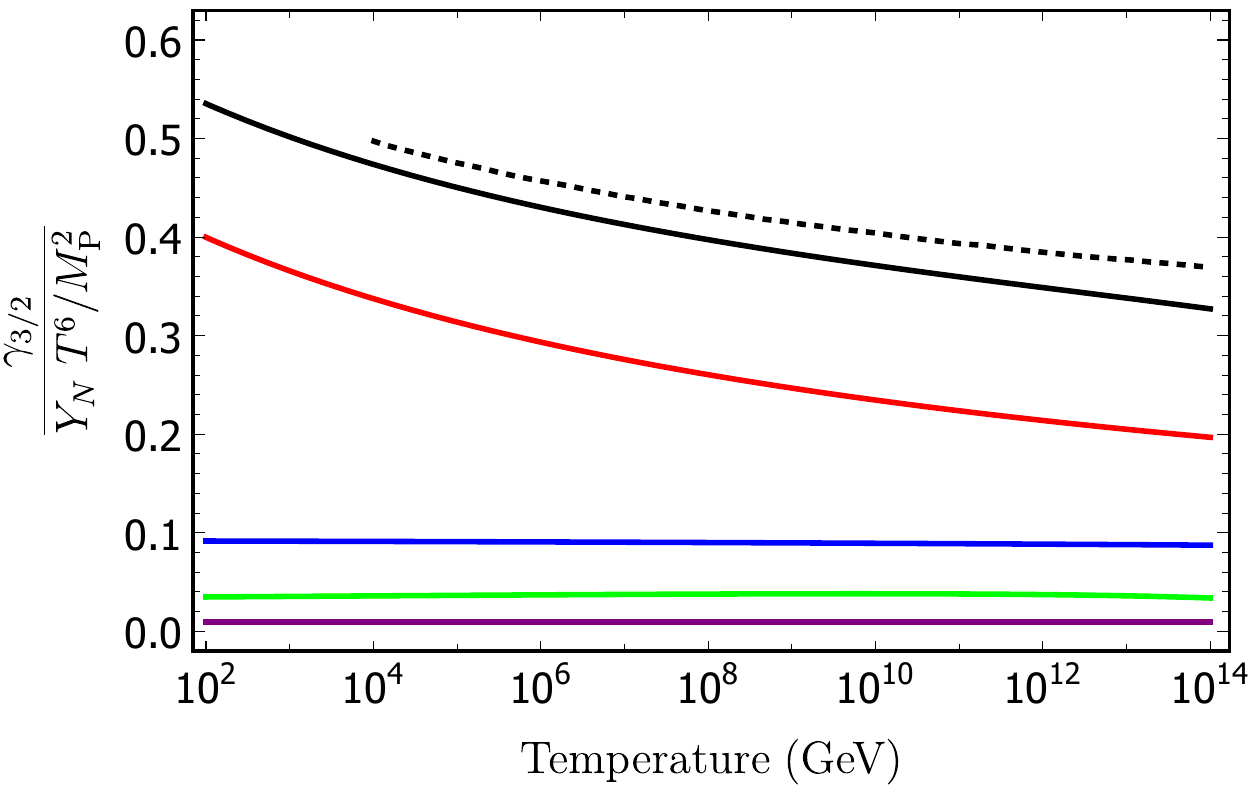}
\caption{The gravitino production rates divided by $Y_N\, T^6  / \mplanck ^2$. The solid curves represent in order
the total rate (black) given by~\eqref{gamma}, the $SU(3)_c$ (red), $SU(2)_L$ (blue), and  $U(1)_Y$ (green) rates 
given by~\eqref{gamma:paramtrization}, and the top Yukawa rate (purple) given by~\eqref{gamma:top}. The upper dashed
 curve is the total production rate obtained in~\cite{Rychkov:2007uq}. The top Yukawa coupling $\lambda _t$ has 
 been taken equal to 0.7 so that our result can be directly compared with that in~\cite{Rychkov:2007uq}.} 
\label{fig:gamma}
\end{figure}
where the constants $c_N$ and $k_N$ depend on the gauge group and their values are given in Table~\ref{table:2}. 
In Fig.~\ref{fig:gamma} we summarize our numerical results for the gravitino production rates divided by $Y_N\, T^6  / \mplanck ^2$. 
Especially, for the case of the top Yukawa contribution,  in $Y_N$ the $m^2_{\lambda_N}$ has to be replaced by $ A_t^2 $.
The colored solid curves represent the $SU(3)_c$ (red), $SU(2)_L$ (blue), and  $U(1)_Y$ (green) rates given by~\eqref{gamma:paramtrization} 
and the top Yukawa rate (purple) given by~\eqref{gamma:top}, while the black solid curve is the total result given by~\eqref{gamma}. 
The dashed black curve  corresponds  to the  total result from~\cite{Rychkov:2007uq}. 
For the sake of comparison, we have also  chosen $\lambda _t = 0.7$~\cite{foot5}.   

Despite  the analytical and numerical discrepancies with~\cite{Rychkov:2007uq},  it is  interesting   
that our result for the total gravitino production rate is only $5\%-11\% $ smaller  than  that in~\cite{Rychkov:2007uq}.  
Being unable to explain   this  quantitively in details, we assume that the aforementioned 
differences have opposing effects on the total result.
For convenience, in Fig.~\ref{fig:gamma},  universal gauge coupling 
unification  is assumed at the grand unification (GUT) scale $\simeq  2\times 10^{16}\, \GeV$, but 
certainly the result in~\eqref{gamma:paramtrization} can be used independently of this assumption.
Equation~\eqref{gamma:paramtrization} along with the numbers in Table~\ref{table:2} 
is  the main result of this paper. 
\section{The Gravitino abundance}
The Boltzmann equation for the gravitino number density $n_{\tiny 3/2}$ is 
\beq
\dot{n}_{\tiny 3/2} + 3H n_{\tiny 3/2} = \gamma_{\tiny 3/2}\,,
\label{eq:boltzmann}
\eeq
where $H$ is the Hubble constant and the dot denotes time differentiation. The gravitino abundance  is defined as
\beq
Y_{\tiny 3/2}= {n_{\tiny 3/2} \over n_{\rm rad} }\,,
\label{eq:abundance}
\eeq
with $n_{\rm rad}=\zeta(3)T^3/\pi^2$. Substituting~\eqref{eq:abundance} into~\eqref{eq:boltzmann},  
 we obtain that the gravitino abundance for $T \ll T_{\rm reh}$ is given by~\cite{Ellis:2015jpg}
\beq
Y_{\tiny 3/2}(T)\simeq {\gamma _{\tiny 3/2}(T_{\rm reh}) \over H(T_{\rm reh}) \,\,  n_{\rm rad}(T_{\rm reh}) }\,\, {g_{*s}(T) \over g_{*s}(T_{\rm reh})}\,,
\label{eq:abundance_app}
\eeq
where  $g_{*s}$ are the effective entropy degrees of freedom in the minimal supersymmetric Standard Model (MSSM).
In this, we assumed that inflaton decays are instantaneous, as the Universe is thermalizing. In~\cite{Ellis:2015jpg,Garcia:2017tuj}, the case of not instantaneous inflaton decay has been taken into account. In particular, following~\cite{Garcia:2017tuj}, in the case of gravitino DM, a correction   factor  $\sim 10\%$ is expected for not  instantaneous inflaton decays.
\begin{figure}[t]
\includegraphics[width=0.483\textwidth]{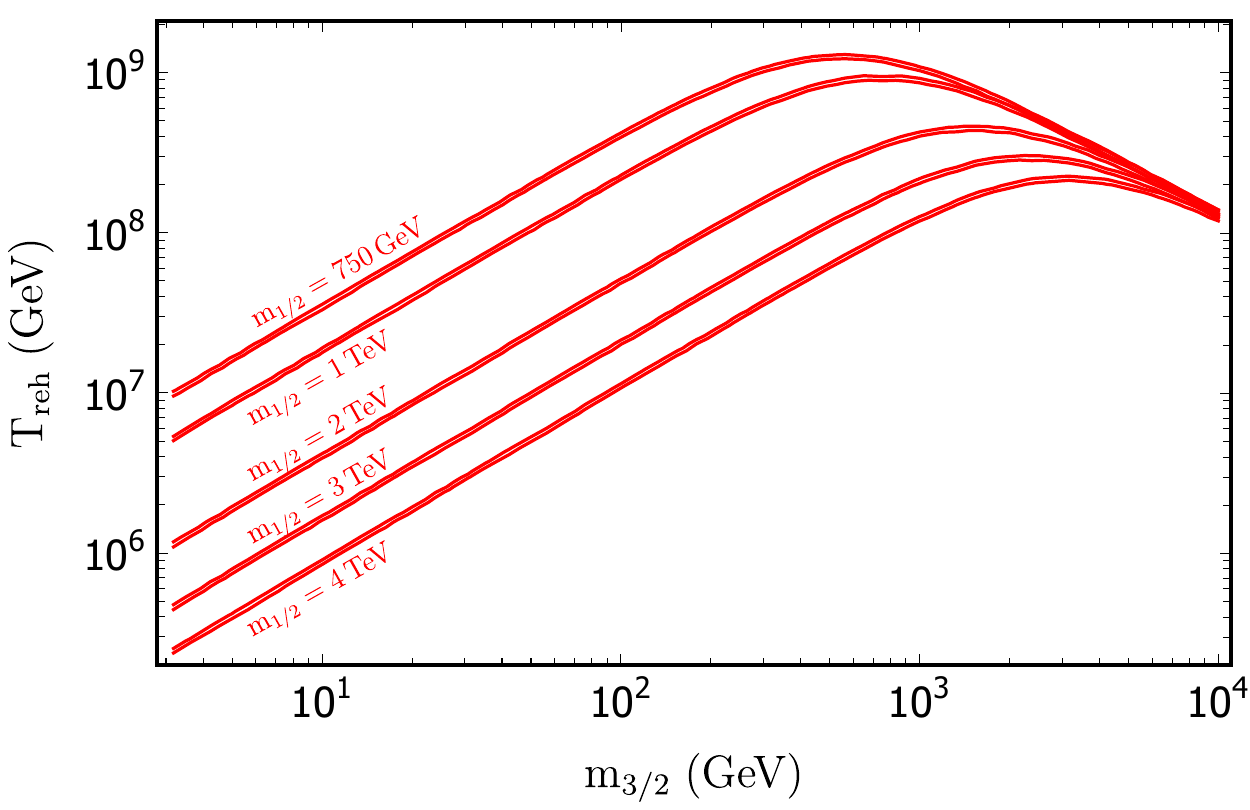}
\caption{The cosmologically accepted   $3 \sigma$ regions for  the gravitino thermal abundance for various
values of $m_{\tiny 1/2}$ between   $750\GeV$ and  $4\TeV$. The trilinear coupling $A_t$ 
has been ignored and the top Yukawa coupling is $\lambda _t =0.7$.}
\label{fig:Treh_m}
\end{figure}

Following the latest data from the Planck satellite, the cosmological accepted value for the 
DM density in the Universe  is
 $\Omega _\mathrm{DM} h^2= 0.1198 \pm 0.0012$~\cite{Aghanim:2018eyx}.
Assuming that the thermal gravitino abundance amounts to the  observed DM,  we obtain that
\bea
\Omega _\mathrm{DM} h^2&= &{\rho_{\tiny 3/2}(t_0)  \,  h^2 \over \rho _{\rm cr} } = {m_{\tiny 3/2} \, Y_{\tiny 3/2}(T_0)\,  n_{\rm rad}(T_0) \,\, h^2 \over \rho _{\rm cr} } \nn
 & \simeq & 1.33\times 10^{24} \,  {m_{\tiny 3/2} \,\, \gamma_{\tiny 3/2}(T_{\rm reh}) \over T_{\rm reh}^5 }\,,
\label{eq:dmdensity}
\eea
where $\rho _{\rm cr}=3\, H_0^2 \mplanck ^2$ is the critical energy density,  $H_0=100\, h\, {\rm km/(s\, Mpc)}$ is
 the Hubble constant,  and $T_0 =2.725\,K$   the 
 cosmic microwave background  temperature   today.  The entropy degrees of freedom at 
the associated temperatures are $g_{*s}(T_0)=43/11$ and $g_{*s}(T_{\rm reh})=915/4$. 
The last  number equals  to the effective energy degrees of freedom  for $H(T_{\rm reh})$   in the MSSM too. 
Figure~\ref{fig:Treh_m} illustrates the $3\sigma$  regions resulting from~\eqref{eq:dmdensity}
for various values of $m_{\tiny 1/2}$.
In this figure the trilinear coupling $A_t$ has been ignored and the top Yukawa coupling is $\lambda _t =0.7$,  as previously.
As before,  gauge coupling  unification  is assumed, as well as a universal gaugino mass $m_{\tiny 1/2}$ at the GUT scale.

For large gravitino mass, the reheating temperature is $m_{\tiny 1/2}$ independent, as the 
characteristic factor $m_{\lambda_N}^2 / (3 m^2_{\tiny 3/2})$ becomes negligible for $m_{\tiny 1/2} \ll m_{\tiny 3/2}$. 
Assuming that $m_{\tiny 1/2} \gtrsim\textit{} 750 \, \GeV$,   as it is suggested 
by the recent LHC data~\cite{Aaboud:2017hrg,Sirunyan:2019ctn} 
on  gluino searches,   from Fig.~\ref{fig:Treh_m}  we infer that for maximum 
$T_\mathrm{reh}\simeq  10^9\, \GeV$ the corresponding gravitino mass is  $m_{\tiny 3/2} \simeq 550\, \GeV $. 
Allowing for a reheating temperature an order of magnitude smaller,  $T_\mathrm{reh} \simeq  10^8 \GeV$, for
the same gravitino mass, $m_{\tiny 1/2}$ can go up to $3-4 \TeV$. 
\section{ Conclusions}
In this paper, we have calculated the gravitino thermal abundance, using 
the full one-loop thermally corrected  gravitino self-energy. Having rectified  the main analytical 
formulas  for the gravitino production  rate,  we  have computed it numerically without approximation.
 We offer a simple and useful parametrization of our final result.
In the context of minimal supergravity models, assuming gaugino mass unification, we have updated the bounds on  the
 reheating temperature for certain gravitino masses. In particular,  saturating the current  LHC  gluino mass
  limit $m_{\tilde g} \gtrsim 2100 \GeV$, we find that  a  maximum reheating temperature 
$T_\mathrm{reh} \simeq  10^9 \GeV$  is compatible to  a gravitino mass $m_{\tiny 3/2}  \simeq 500 - 600\, \GeV $. 

It should be noted that trying to constrain the reheating temperature  by  applying  
 the cosmological  data on   gravitino DM scenarios illuminates us whether  thermal leptogenesis is a possible  mechanism for generating baryon asymmetry or not. Successful  thermal leptogenesis requires high temperature,
 $T_\mathrm{reh} \gtrsim 2\times  10^9\, \GeV$~\cite{Giudice:2003jh,Antusch:2006gy,Buchmuller:2005eh}, which is marginally bigger than the maximum  reheating temperature obtained in our model using the lowest $m_{\tiny 1/2}  $ mass demonstrated in the  recent  LHC  data~\cite{Aaboud:2017hrg,Sirunyan:2019ctn}.  
In any case, there are many alternative models  for baryogenesis. 
In addition, as it has been pointed out before, the thermal gravitino abundance is in general   a part of the 
whole DM density and the inclusion of  other  components will affect  the  phenomenological analysis.
\begin{acknowledgments}
 The authors thank  A. Lahanas and J. Pradler for   useful discussions.
 This research is co-financed by Greece and the European Union (European Social Fund- ESF) through the 
 Operational Programme \textquote{Human Resources Development, Education and Lifelong Learning} 
 in the context of the project \textquote{Strengthening Human Resources Research Potential
  via Doctorate Research - 2nd Cycle} (MIS-5000432), implemented by the State Scholarships Foundation (IKY). 
   This research work was supported by the Hellenic Foundation for Research 
   and Innovation (H.F.R.I.) under the ``First Call for H.F.R.I. Research Projects to support 
   Faculty members and Researchers and the procurement of high-cost research equipment grant'' (Project Number: 824).
 \end{acknowledgments}



\begin{thebibliography}{99}
\bibitem{foot1}
As it was argued in~\cite{Bolz:2000xi} and we have checked 
numerically,   the effect of taking into account the statistical factor $f_c$ can be around
$-10$\%  ($+20$\%) if $c$ is a fermion (boson).
\bibitem{foot4}
In principle, we use the $\d$-functions to reduce the number of the phase-space integrations. Further details will appear in~\cite{Eberl:20020gr}.
\end{thebibliography}
\end{document}